\documentstyle[11pt,psfig]{article}
\def\upcite#1{$^{\cite{#1}}$}
\def\bfr{{\bf r}}
\def\bfv{{\bf v}}

\def\vev#1{{\left\langle #1 \right\rangle}}
\def\iras{{\it IRAS\/}}
\def\etal{{\it et al.\/}}
\def\kms{\ifmmode {\rm \ km \ s^{-1}}\else $\rm\, km \ s^{-1}$\fi}
\def\fig #1, #2, #3 {
\smallskip
\centerline{\psfig{figure=#1,height=#2 in,width=#3 in}}
}
\begin{document}
\title{Large-Scale Structure in the Distribution of Galaxies as a
Probe of Cosmological Models} 
\author{Michael A. Strauss\footnote{Alfred P.\ Sloan Foundation
Fellow}{$\,\,$}$^,${$\,\!$}\footnote{Cottrell Scholar of Research
Corporation} , Princeton University Observatory, Princeton, NJ
08544\\
\tt (strauss@astro.princeton.edu)} \maketitle 

{\bf The last 20 years have seen an explosion in our understanding of
the large-scale distribution and motions of galaxies in the nearby
universe.  The field has moved from a largely qualitative,
morphological description of the structures seen in the galaxy
distribution, to a rich and increasingly rigorous statistical
description, which allows us to constrain cosmological models.  New
surveys just now getting underway will be unprecedented in their
uniformity and volume surveyed.  The study of the evolution of
large-scale structure with time is now becoming feasible.}

\section{Introduction}
\label{sec:intro}

In 1970, Alan Sandage wrote an article\upcite{Sandage70} describing
observational cosmology as a ``search for two numbers'', namely the
Hubble Constant $H_0$, which sets the overall scale of the universe,
and the acceleration parameter $q_0$, which measures its
curvature\upcite{Weinberg72,Peebles93}.  Although the values of these
two parameters are still very much a matter of
contention\upcite{DialogueMeeting}, the field of observational
cosmology has broadened considerably since then, as we have become
aware of the richness of the information encoded in the large-scale
distribution of galaxies.  The observed
distribution of galaxies on the sky shows hints of structure on large
scales\upcite{deVaucouleurs53,Peebles77}, but without distance 
information to each individual galaxy, one is only seeing the galaxy
distribution in projection.  However, Hubble's
law\upcite{Hubble36} states that due to the expansion of the universe,
the redshift of a galaxy $cz$ is proportional to its distance $r$: 
$cz = H_0 r$.  Thus the measurement of the redshifts of galaxies allows
their distances to be inferred, yielding the full three-dimensional
distribution of galaxies.  In the late 1970's and early
1980's, improvements in spectrographs and detector technology allowed
redshifts of large numbers of galaxies to be measured efficiently and
quickly, and the first large redshift surveys of galaxies were carried
out\upcite{Gregory78,Huchra83,Yahil80,KOSS81,Bean83}, many of them at
Kitt Peak.

These early redshift surveys showed large voids and walls in the
galaxy distribution\upcite{Gregory78,KOSS81,Davis81}, but the richness
of the structure present really only hit
home for the majority of astronomers with the publication of the
first declination slice of the Center for Astrophysics redshift
survey\upcite{deLapparent86} (Figure 1). 
Galaxies are distributed on the
walls of huge voids, as large as 100 Megaparsec (Mpc,
where we assume a Hubble constant of $100 \kms\,\rm Mpc^{-1}$
throughout) in extent.  A coherent
structure is seen spanning the entire map (``The Great
Wall''\upcite{Geller89}), causing us to ask whether yet larger
structures would become apparent with a redshift survey covering a yet
larger volume\upcite{deLapparent88}.  Figure~1 is a
two-dimensional slice through the galaxy distribution (albeit not
suffering from the projection effects seen in the distribution of
galaxies on the sky without redshifts).  The extension of this survey in
the third dimension\upcite{Geller89} has shown that galaxies are
indeed roughly distributed on the walls of spherical voids, as Figure
1 was originally interpreted, although the
galaxy distribution is somewhat more filamentary than this
mental picture suggests\upcite{Vogeley94}.  This has been confirmed
further with the Las Campanas Redshift Survey\upcite{Shectman96}; with
over 20,000 galaxies, it is the largest single redshift survey to
date. 

In any case, a map like this broadens the range of questions which we
want to ask in the field of observational cosmology.  In addition to
the values of $H_0$ and $q_0$, we would like to know:
\begin{itemize} 
\item What is the topology of the galaxy distribution on
various scales?  What are the largest coherent structures that exist
in the galaxy distribution?  The Cosmological
Principle\upcite{Peebles93} states that the universe is homogeneous
and isotropic on the largest scales; is this indeed observed? 
\item How did this structure form in the first place?  How might we
constrain the parameters describing the expanding universe, and models
for structure formation with these data?  What is the 
connection to the formation of galaxies themselves?  And how does the
inhomogeneous structure relate to the {\em peculiar velocities},
galaxy motions above and beyond the Hubble flow? 
\item Galaxies and clusters of galaxies are gravitationally dominated
by dark matter\upcite{Kormendy87}; indeed, only of order 1\% of the
matter in the universe is directly visible.  What is the distribution of the
galaxies relative to the dark matter; are the galaxies a biased tracer
of the mass distribution of the universe? What is the nature of
the dark matter which dominates the gravity, and therefore the
dynamics, of the universe?
\end{itemize}

We can now begin to address these questions with
observations of the distribution and motions of nearby galaxies, but
we are very much limited by the non-uniformity and finite volume
probed by existing datasets.  As we describe below, we do most of our
work in the context of a standard model whereby structure forms via
gravitational instability from tiny initial fluctuations, but with
present data, this paradigm is not properly tested, and its parameters
are only poorly known.  New surveys now getting underway, both
at low and high redshift, should allow us to address all of these
questions in much more quantitative detail than has been possible in
the past. 

\section{Quantifying the Distribution of Galaxies}

The rms 
fluctuations of the observed galaxy density field are very large on
small scales, of order unity within spheres of radius 8 Mpc, dropping
as a power law with scale, becoming a few percent at several 
tens of Mpc.   It therefore makes
sense to reference the density field of galaxies to its mean.
Let $\rho(\bfr)$ be the observed galaxy density field smoothed with
some uniform kernel; the density {\em fluctuation} field is defined as
$\delta(\bfr) \equiv (\rho(\bfr) - \vev{\rho})/\vev{\rho}$.  This
quantity can be measured from flux-limited redshift surveys, after suitable
accounting for the fall-off in observed galaxy number density with
distance due to the flux limit\upcite{SW}.  Figure~2
shows the observed galaxy distribution of a redshift
survey of galaxies\upcite{Fisher95} from the database of the {\it
Infrared Astronomical Satellite} (\iras), in the Supergalactic
Plane, in which many of the more prominent structures in
the Local Universe lie.  

We
interpret these observations in the context of the 
prevailing Big-Bang model of 
cosmology\upcite{Peebles93,Coles96,Padmanabhan95,KolbTurner90}.
Observations of the Cosmic Microwave Background
(CMB)\upcite{Partridge96} show that at a redshift $z \approx 1100$,
the universe showed deviations from uniformity at one part in
$10^5$ on scales of several hundred Mpc\upcite{White94}.  In standard
inflationary models for 
the Big Bang\upcite{KolbTurner90}, these initial fluctuations have a
Gaussian one-point distribution, and the Fourier modes into which they
might be decomposed have random phases.  To the extent that this is
true, the {\em power spectrum} of the density field, the mean square
amplitude of the Fourier mode as a function of wavenumber $k$, is a
complete statistical description of the
fluctuations\upcite{Bertschinger96}.  The process of gravitational
instability gives rise to the structures that we see, and indeed the
numbers work out; {\it a priori\/} predictions of the amplitude of CMB
fluctuations from the observed distribution of
galaxies\upcite{Peebles82}, which involve a large extrapolation both in
time and in spatial scale, were indeed of the order of $10^{-5}$, as
observed.  

In linear perturbation theory, while the amplitude of the power
spectrum grows with time, the shape of the power spectrum does
not evolve.   Therefore the 
measurement of the present-day power spectrum allows us to determine
the shape of the initial power spectrum, which encodes in it
information on the Cosmological Density Parameter $\Omega$ and the
nature of the dark matter\upcite{White94}.
Measurements of the present-day galaxy power spectrum, and its Fourier
Transform, the correlation function, from both angular and redshift surveys of
galaxies\upcite{Peacock97} have been instrumental in demonstrating that the
standard $\Omega = 1$ Cold Dark Matter model overpredicts the
observed power on small scales\upcite{Fisher93,Ostriker93}, thereby
ruling out this once-standard model.  However, the quantitative
agreement between the power spectra found from different redshift
surveys is not 
particularly good, and their interpretation in terms of theoretical
initial power spectra has been difficult.  The two principal reasons
for this difficulty are the presence
of peculiar velocities, and our ignorance of the relative distribution
of galaxies and dark matter.

Look again at Figure~1.  As described in the figure caption, the
redshift distribution of galaxies is distorted by peculiar
velocities.  The radial component of these 
peculiar velocities $\bfv$ modify Hubble's law, 
\begin{equation} 
cz = H_0 r + \hat \bfr\cdot \left[\bfv(\bfr) - \bfv({\bf
0})\right],
\label{eq:redshift-distance}
\end{equation}
where $\bfv({\bf 0})$ is the peculiar velocity of the Local Group.
The net effect is to take a dense 
concentration of galaxies and spread it out in redshift space; the
clustering therefore appears weaker than it would in real space.  Thus
on the relatively small scales of clusters, peculiar velocities cause
the power spectrum to be {\em underestimated}. 

  On large scales, the opposite effect happens\upcite{Kaiser87}.  A
large overdense region of space gravitationally attracts the galaxies
that lie near it, giving them peculiar velocities that make them
appear {\em closer} to the overdensity in redshift space than in real
space.  The net effect is that the structure appears {\em compressed},
and therefore of higher 
density, in redshift space than it really is.   This effect can be
measured directly via the anisotropy this effect induces in the
clustering, 
although current measurements are very much limited by the
finite volumes surveyed thus far\upcite{Hamilton98}.
Practical methods are needed to fit redshift survey data
for the underlying power spectrum, accounting for peculiar velocities
both in the linear and non-linear regime. 

  Even more vexing a problem is the relative distribution of galaxies
and dark matter.  Astronomers have long parameterized their ignorance
of this issue by hypothesizing a direct proportionality between the
density fluctuations of galaxies $\delta_g$ and dark matter $\delta_{DM}$,
$\delta_g= b \delta_{DM}$, where the {\em
biasing parameter} $b$ is 
independent of location or smoothing
scale\upcite{Kaiser84,DekelRees87}.  Although there are plausibility
arguments that this might be the
case\upcite{Kauffman98}, it is quite  
difficult to test this directly.  Cosmological simulations show that
the locations at which galaxies form depend on various physical
quantities in addition to the local dark matter density.  Thus the
relationship between the galaxy and mass distribution is both
non-linear and stochastic, and differs from
one sample of galaxies to another.  People are only now starting
to investigate how this propagates to cosmological inferences drawn
from the observed distribution of galaxies\upcite{DekelLahav98}.

One important way to get a handle on this is to measure the
large-scale distribution of galaxies of different types.  The cores of
rich clusters of galaxies are almost entirely composed of elliptical
and lenticular galaxies\upcite{Hubble36,Dressler80}, while these two
populations represent only 20\% of galaxies in the field.  If the
distribution of different types of galaxies do not agree with one
other, they cannot all agree with the distribution of dark matter!  On
scales larger than clusters, it is known that there are not {\em gross}
differences in the relative distribution of galaxies of different
types or different luminosities\upcite{Huchra91}, although the
clustering strength does depend on these
quantities\upcite{Davis76,Davis88,Guzzo97}.  These analyses are 
limited by the small samples and poor morphological information
available; with larger and better samples, we might
hope to learn much more about the nature of bias, which, as hinted at
above, may teach us about the process of galaxy formation. 

Even if the initial density field is perfectly Gaussian, and thus can
be completely described by the power spectrum, gravitational
instability theory predicts that as the fluctuations 
grow in strength, the 
distribution of $\delta$ must develop a skewness, whose value can be
calculated to leading order in perturbation
theory\upcite{Juszkiewicz93}.  In addition, there are 
classes of models in which large-scale clustering is seeded by exotic
initial structures such as textures\upcite{Spergel93}, which imprint
non-Gaussian features in the initial density field.  Thus measurements
of the non-Gaussianity of the galaxy distribution allow one to test
gravitational instability theory, and to look for signatures of these
seeds.  Alternatively, the topology of the galaxy distribution is a
measure of the correlation of the relative phases of Fourier modes.
One measures the genus number of surfaces of constant galaxy density
\upcite{Gott86}; comparison with the prediction in the random phase
case are a clean measure of non-Gaussianity.
Measurements of high-order correlations from redshift
surveys\upcite{Bouchet93,KimStrauss98} and angular
catalogs\upcite{Gaztanaga95} are in impressive accord with
gravitational instability predictions.  Similarly, topology
measurements\upcite{Vogeley94} show the qualitative effects
expected from non-linear growth of fluctuations from Gaussian initial
conditions.  Indeed, there is as yet no convincing evidence for
initial non-Gaussian seeding of the density field.  However, existing
data are not powerful enough to make these constraints very strong.

\section{Peculiar Velocities}

We have seen that large-scale inhomogeneities in the galaxy
distribution should give rise to peculiar velocities. 
In linear perturbation theory, there is a linear relation
between the density and velocity fields: $\nabla \cdot \bfv(\bfr) = -
H_0 \Omega^{0.6} \delta_{DM}(\bfr)$\upcite{Peebles93}.  The Hubble
Constant $H_0$ is 
identically equal to unity if one measures the distances to galaxies
in redshift units.  Thus a comparison between the galaxy density
distribution and the peculiar velocity field allows a measurement of
the Cosmological Density Parameter $\Omega$.  However, because
peculiar velocities are caused by the gravitational attraction of all
matter, while the density fluctuations we can measure directly are
those of galaxies, we are affected by biasing in this comparison.  In
particular, for linear biasing, the 
relationship between the galaxy velocity and density fields becomes 
\begin{equation} 
\nabla \cdot \bfv(\bfr) = - \beta \delta_g(\bfr),
\label{eq:del-dot-v} 
\end{equation}
where $\beta \equiv \Omega^{0.6}/b$.  

By far the best-measured peculiar velocity is our own; we measure a
0.1\% dipole moment in the temperature of the CMB\upcite{Partridge96}, which
is interpreted as a Doppler effect due to the Sun's motion relative to
the rest frame defined by all the material which
radiates the CMB photons we detect.  Correcting for the rotation of the Milky
Way, and the infall of the Milky Way to the barycenter of the Local
Group yields a peculiar velocity of 620 \kms\ towards Galactic
coordinates $l = 276^\circ, b = +30^\circ$.  Using an integral form of
equation~(\ref{eq:del-dot-v}) allows this value to be compared with
the prediction from a redshift survey; a detailed
analysis\upcite{Strauss92} yields $\beta = 0.55^{+0.20}_{-0.12}$.

The radial component of the peculiar velocity for a galaxy
follows immediately from independent measures of its distance and
redshift (equation~\ref{eq:redshift-distance}, where as before, we
work in units in which $H_0 \equiv 1$).  
One determines the relative distances of galaxies using {\em distance 
indicator relations}\upcite{SW}, whereby their rotation
velocities (spirals) or internal velocity dispersions (ellipticals)
are related to their optical luminosities; a measurement of their
apparent brightnesses then yields their distances via the inverse
square law.  The telescopes at Kitt Peak have been instrumental for
the development and refinement of these distance indicators, and
indeed, much of the data in the analyses described below was obtained
at Kitt Peak. 

These distance indicator relations show appreciable scatter, allowing
the measurement of the distances of individual galaxies to 
15-20\% accuracy.  For a galaxy at a distance of 40 Mpc, this is a
600 \kms\ error at best, which of course propagates directly into the
inferred peculiar velocity.  As this is of the order of the peculiar
velocity itself, the signal-to-noise ratio in the inferred peculiar
velocity field is less than one per galaxy.  Moreover, the results are
subject to substantial statistical biases depending on the details of
how the distance indicator relation is used\upcite{SW,Teerikorpi97}.
One therefore requires a heavy grouping or smoothing algorithm, or an
elaborate statistical method, to extract useful information from such
data.

In order to trace out the full
velocity field, we use peculiar velocity samples that cover much of
the sky\upcite{Willick97,daCosta97}. 
A number of papers\upcite{daCosta97,Shaya95,DNW,VELMOD2}
have developed rigorous techniques for comparing peculiar velocity and
redshift survey data via the integral form of
equation~(\ref{eq:del-dot-v}).  These results 
indicate the likelihood of systematic effects 
at some level in the TF data, probably due to the difficulty of
matching separate samples across the sky\upcite{Willick97}; when these
are corrected for, authors agree that the data are consistent with
equation~(\ref{eq:del-dot-v}), and that $\beta = 0.5-0.6$
(ref.~\cite{Shaya95} is the exception, with the smaller value of
$\beta = 0.34 \pm 0.13$).  

Alternatively, one can work from
equation~(\ref{eq:del-dot-v}) in its differential form\upcite{Dekel94}.
To the extent that the peculiar velocity field is irrotational (as
expected under gravitational instability on all but the smallest, most
non-linear scales), it can be expressed as the gradient of a scalar
potential: $\bfv = -\nabla \Phi_v$.  This potential field can be
determined by integrating the observed {\em radial} velocity field
$\hat \bfr \cdot \bfv(\bfr)$ (after suitable smoothing) along radial
rays.  The full three-dimensional velocity field \bfv(\bfr) is the
gradient of the resulting field, from which the divergence, or a
suitable non-linear extension\upcite{Nusser91}, can be measured.  The
effective smoothing of the resulting map is set by the initial
smoothing of the raw peculiar velocities; in practice, this smoothing
is a 1200 \kms\ Gaussian.

  A comparison of the resulting inferred density field with the \iras\
galaxy density field\upcite{Sigad98} shows beautiful consistency, as
predicted from linear biasing and gravitational instability theory.
This analysis finds $\beta = 0.89 \pm 0.12$, in strong {\em
disagreement} with the analyses quoted thus far (some of which used
essentially the same redshift survey and peculiar velocity data!).
The reason for this disagreement is not yet clear.  This may be due to
subtle systematic effects in the data; alternatively, it is possible
that our assumption of a scale-independent bias is incorrect, although
a self-consistent model which explains all the data has not yet been
developed. 

Although we do not have a direct measure of the value of the bias
parameter $b$, current theoretical models indicate that is of the
order of unity.  Thus the results above translate into values of
$\Omega$ around 0.3 and 1.0, respectively.  The former value is close
to what is inferred from studies of the abundances and evolution of
clusters of galaxies\upcite{Dekel97,Bahcall98}, and the observed shape of
power spectrum (see the previous section), while the latter is
the ``cleanest'' prediction of inflationary models.  The community has
recently moved over towards the $\Omega \approx 0.3$ camp for the most
part; the analysis cited above giving $\beta = 0.89$ remains one of
the strongest arguments for an appreciably larger value of $\Omega$.
But the result will not be settled until it is understood why different
analyses with the same data are giving such disparate results.

\section{The Future}

The various quantitative analyses of large-scale structure described
above are limited by systematic effects, either due to
non-uniformities in the data at some level, or the smallness of the
volumes surveyed.  For example, consider the measurement of the galaxy
power spectrum on the largest scales.  This is a crucial quantity, as
it can be a strong constraint on cosmological models.  Systematic
errors in the photometry of galaxies from which a sample is selected
for a redshift survey will mimic large-scale gradients in the density
field, causing systematic errors in the inferred power spectrum.  In
addition, on scales approaching the scale of the sample, one by
definition only has a small number of independent volumes on which to
measure density fluctuations.  Thus for any redshift survey, there
will be a scale, comparable to the volume of the survey, for which the
sample is not fair for measuring the power spectrum.

  These considerations motivate us to consider the largest, most
uniform possible survey of the galaxy distribution.  The Sloan Digital
Sky Survey\upcite{GunnWeinberg95,Knapp97} is such a survey.
  Over the course of five years, a wide-field CCD camera on a
dedicated wide-field 2.5m telescope will
survey the entire Northern Galactic cap (1/4 of the Celestial Sphere)
in 5 colors 
to a depth of $r' = 23.1$, creating a highly uniform digital catalog
of over $10^8$ galaxies and stars.  The brightest million
of these galaxies will be targets for a redshift survey, to be carried
out by a pair of fiber-fed multi-object double spectrographs on the
same telescope.  The resulting database will be an order of magnitude
increase in the volume probed by redshift surveys, and a factor of 40
increase in the number of galaxies in a single 
survey.  More important, it will have a much tighter control of
systematic errors than previous surveys.  Finally, the detailed
photometric data available for each galaxy will allow 
large-scale structure studies to be done as a function of galaxy
morphological type, color, and luminosity, allowing us to quantify the
relative bias, and to develop more complete models for the underlying
dark matter density field than the simplistic linear bias assumption
allows us. 

A competing survey\upcite{Colless} uses a wide-field spectrograph
attached to the Anglo-Australian Telescope to obtain redshifts for
250,000 galaxies selected from photographic plates.  This survey has
obtained redshifts for almost 10,000 galaxies as of this writing.  As
their galaxy selection is based on shallower, single-color
photographic plate material, their control on systematic effects will
not be as great as in the Sloan survey.

  These two surveys will make definitive measurements of the nature of
large-scale structure in the nearby universe.  Analyses show that they
should measure the power spectrum to an accuracy sufficient to measure
a large host of cosmological parameters, especially when coupled with
measurements of the CMB\upcite{Wang98}.   

A new frontier in the field is just now opening up; the study of
large-scale structure at high redshift. The finite speed of light
means that we see distant galaxies at a time when the universe was
younger than it is today.  Thus we can study directly the growth of
large-scale structure with time, by measuring the distribution of
distant galaxies.  Indeed, gravitational instability theory makes
definite predictions for the evolution of structure with redshift,
which can be checked from such measurements.

  One probe of the evolution of clustering has been via wide-field
photometric surveys of faint galaxies.  
The angular correlations of galaxies can be measured directly from deep
images of the sky\upcite{Efstathiou91b,Infante94,Lauer98}; some of the
best work 
in this field is being done at Kitt Peak.  Indeed, techniques
have been developed recently for measuring approximate redshifts of
galaxies from their broadband colors\upcite{Hogg98}, which allow the
redshift dependence of the angular clustering to be measured
directly\upcite{Connolly98}.  The KPNO 4-m has been instrumental in
obtaining the broadband photometric data one needs to calibrate and
measure these photometric redshifts.

With accurate spectroscopic redshifts of high-redshift samples, more
detailed analyses are possible.  Two Canadian and French
groups\upcite{CFRS,CNOC2} have carried out extensive redshift surveys
of up to several thousand galaxies in clusters and the field, to a
redshift of $z \approx 1$, and have been able to measure the evolution
of the small-scale clustering of galaxies.  Most dramatically, Steidel
and colleagues\upcite{Steidel96} have been able to identify galaxies
likely to lie at $z > 3$ by their particularly red $U-B$ colors due to
Lyman continuum absorption redshifted to the $U$ band (a crude form of
photometric redshift); spectroscopic surveys of these extremely faint
objects with the 10-m Keck telescopes on Mauna Kea have resulted in
redshifts of literally hundreds of these galaxies.  They have found
astonishingly strong clustering at $z \approx 3$\upcite{Steidel98}.
Gravitational instability theory predicts appreciably weaker
clustering at high redshift for the dark matter; the interpretation,
then, is that this early population of galaxies is much more strongly
biased relative to the dark matter than are galaxies today.  In
retrospect, this indeed makes sense; one generically expects the bias
factor to approach unity as galaxies are gravitationally drawn to the
dark matter potential wells\upcite{Fry96,TegmarkPeebles98}, but it
tells us that the study of the evolution of large-scale structure will
require a deep understanding of the evolution of bias, and therefore
probably also of galaxy formation.  New extensive redshift surveys of
faint galaxies planned for the Keck telescope and other 8-10 meter
class telescopes coming on line around the world\upcite{BigTelescopes}
promise that insights to this problem will be coming soon; the
bottleneck for this work, ironically, is the lack of deep imaging
surveys of galaxies on wide-field 2.5-m class telescopes from which
the objects for the redshift surveys can be selected. Kitt Peak is
leading the world-wide effort to fill this gap (Jannuzi, private
communication). 

The field of large-scale structure is reaching a new point of
maturity, which will allow us to address the big questions in the
field outlined at the end of \S~\ref{sec:intro}.  With the new surveys of
the local universe which are just getting underway, we should have
accurate measurements of a number of cosmological parameters, and a
firm zeropointing of the present-day large-scale structure with which
to compare the exciting results from observations of structure at high
redshift.  These surveys will be large enough to make a definitive
measurement of the topology of the galaxy distribution on the largest
scales.  To the extent that we fit the data convincingly with a
cosmological model without large numbers of ad-hoc parameters, we can
claim to have a complete model for the formation of
structure.  Measurement of large-scale structure as traced by galaxies 
of different types will give us insight into the increasingly
complicated problem of the relative distribution of galaxies and dark
matter.  A full picture will become clearer with observations of the
evolution of galaxy clustering with redshift, together with a more
thorough theoretical understanding of the evolution of bias.

We will really only be able to claim that we have a coherent
cosmological model from these data if the results are consistent with
a variety of other cosmological probes which we have not had space to
cover in this review.  The CMB anisotropies give a complementary
measurement of the mass power spectrum\upcite{White94}; the
Microwave Anisotropy Probe (MAP; see {\tt http://map.gsfc.nasa.gov}),
a satellite which will fly in Fall 2000, is designed to make such a
measurement.  Measurements of quasar absorption lines\upcite{Rauch98}
probe the power spectrum on the smallest scales; recent
work\upcite{Croft98} has shown that the extraction of this information
is quite straightforward.  Large-scale structure can be probed via
the distribution of clusters of galaxies\upcite{Bahcall88}, and as we
mentioned briefly above, their mass function and evolution also gives
a sensitive measure of cosmological parameters; new large complete
surveys in the optical\upcite{GunnWeinberg95,Lauer98} and the
X-ray\upcite{Rosati98} will give us an enormous increase in our
understanding of these objects. 

\def\AJ{{\it Astron.~J.}}
\def\ApJ{{\it Astrophys.~J.}}
\def\ApJS{{\it Astrophys.~J.~Suppl.}}
\def\MNRAS{{\it Mon.~Not.~R.~Astron.~Soc.}}
\def\ARAA{{\it Ann.~Rev.~Astron.~Astrophys.}}

\bigskip
MAS acknowledges the support of the Alfred P. Sloan Foundation,
Research Corporation, and NSF grant AST96-16901.  

\vfill\eject
\begin{figure}
\fig 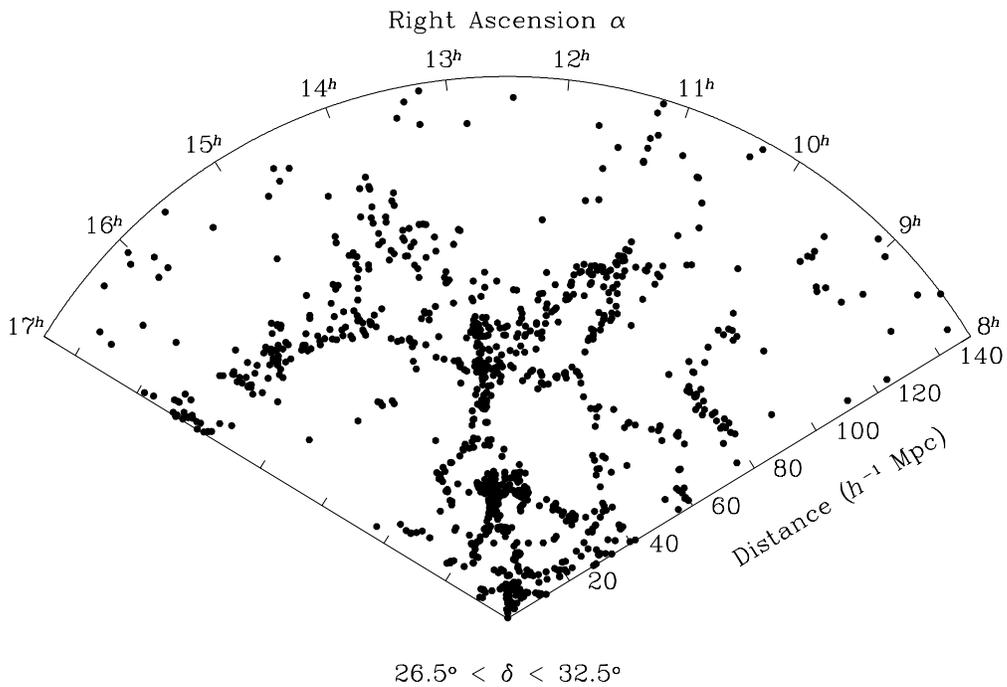, 6, 6
\caption{The distribution of galaxies in redshift space from the survey of
ref.~\cite{Huchra91}.  The sample covers a narrow range of
declination, so redshift is plotted
against right ascension, with declination suppressed.  The large
elongated structure in the middle of the map is the Coma cluster of
galaxies.  In the deep potential well of the cluster, galaxies have
random motions in addition to their motion due to universal expansion,
with a velocity dispersion of order 1000 \kms\upcite{Kent82}. Thus the
redshift diagram of a rich cluster is stretched out 
along the line of sight; the resulting structure is often referred to
as a ``Finger of God'', due to the fact that it points directly to
the observer at the origin. }
\end{figure}

\begin{figure}
\centerline{\psfig{file=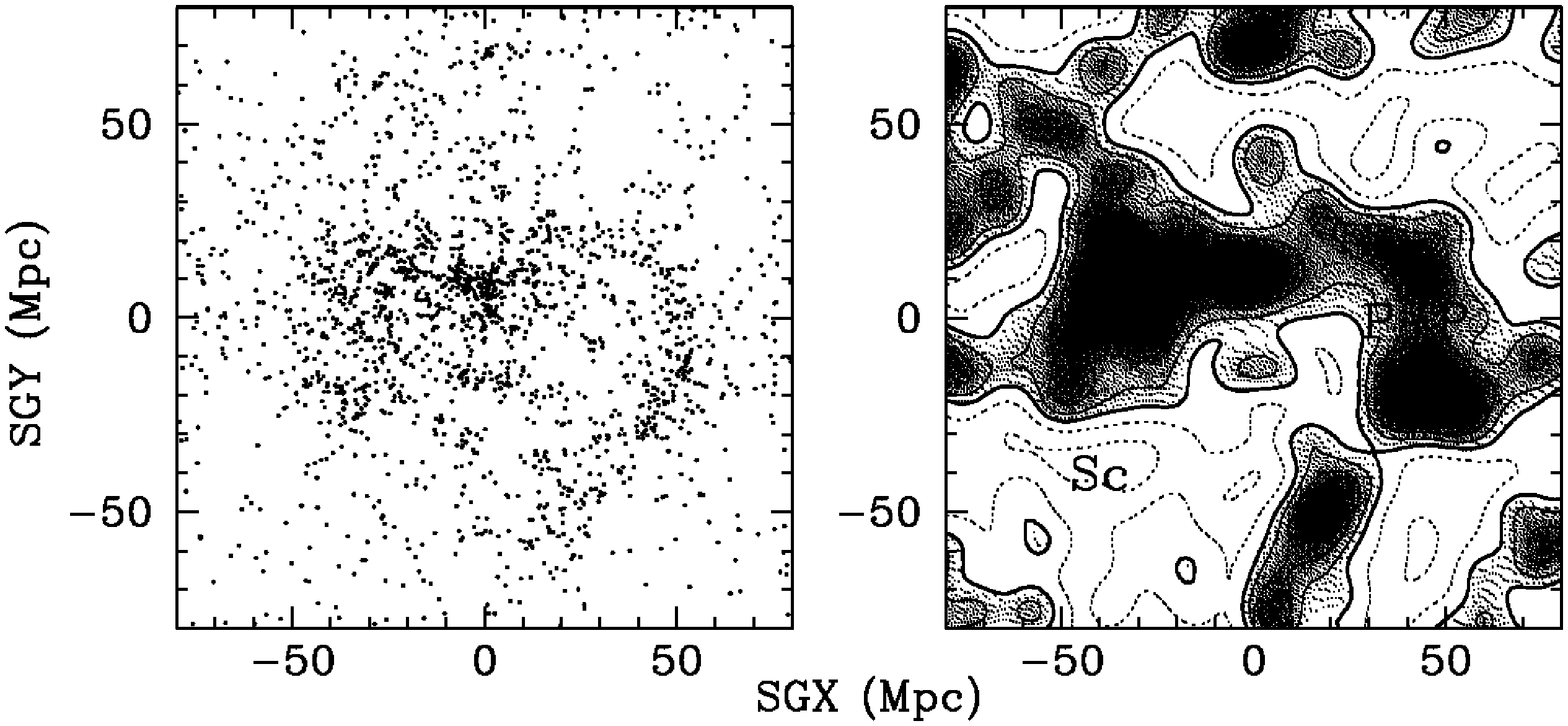,angle=270,width=7in}}
\caption{The distribution of galaxies in redshift space from the survey of
ref.~\cite{Fisher95}.  The left-hand panel shows those galaxies within
$22.5^\circ$ of the Supergalactic plane, while the right-hand panel
shows the corresponding smooth density field, using the methods of
ref.~\cite{Sigad98}.  Labels show several of 
the prominent named structures, including the Virgo cluster (V), the
Perseus-Pisces supercluster (P-P), the Great
Attractor (GA), and the Sculptor Void
(Sc).} 
\end{figure}


\begin{thebibliography}{}
\bibitem{Sandage70} Sandage, A. Cosmology: a search for two
numbers. {\it Physics Today} {\bf 23}, \#2, 34-41
(1970).
\bibitem{Weinberg72} Weinberg, S. {\it Gravitation and Cosmology: Principles
and Applications of the General Theory of Relativity} (New York: Wiley)
(1972).
\bibitem{Peebles93} Peebles, P. J. E. {\it Principles of Physical Cosmology}
(Princeton: Princeton University Press)
(1993).
\bibitem{DialogueMeeting} Turok, N. editor, {\it Critical
Dialogues in Cosmology} (Singapore: World Scientific) 
(1997).
\bibitem{deVaucouleurs53} de Vaucouleurs, G. Evidence for a
local supergalaxy.  \AJ\ {\bf 58}, 30-32
(1953).
\bibitem{Peebles77} Seldner, M., Siebers, B., Groth, E. J., \& Peebles, P. J. E. 
New reduction of the Lick catalog of galaxies. \AJ\ {\bf 84}, 249-256
(1977).
\bibitem{Hubble36} Hubble, E. P. {\it The Realm of the Nebulae} (New Haven: Yale
University Press)
(1936).
\bibitem{Gregory78} Gregory, S. A., \& Thompson, L. A. Low mass
portion of the galaxy clustering spectrum. {\it Nature} {\bf 274}, 450-452
(1978).
\bibitem{Huchra83} Huchra, J. P., Davis, M., Latham, D., \& Tonry,
J. A survey of galaxy redshifts. IV - The data. \ApJS\ {\bf 52}, 89-119
(1983).
\bibitem{Yahil80} Yahil, A., Sandage, A., \& Tammann, G. A. The
velocity field of bright nearby galaxies. III - The distribution in
space of galaxies within 80 megaparsecs - The north galactic density
anomaly. \ApJ\ {\bf 242}, 448-468
(1980).
\bibitem{KOSS81} Kirshner, R. P., Oemler, A., Schechter, P. L., \& Shectman,
S. A. A million cubic Megaparsec void in Bo\"otes? \ApJ\ {\bf 248}, L57-L60
(1981).
\bibitem{Bean83} Bean, A. J., Efstathiou, G. P., Ellis, R. S., Peterson, B. A.,
\& Shanks, T. A complete galaxy redshift sample. I - The
peculiar velocities between galaxy pairs and the mean mass density of
the Universe. \MNRAS\ {\bf 205}, 605-624
(1983).
\bibitem{Davis81} Davis, M., Huchra, J., Latham, D. W., \& Tonry,
J. A survey of galaxy redshifts. II - The large scale space
distribution. \ApJ\ {\bf 253}, 423-445
(1982).
\bibitem{deLapparent86} de Lapparent, V., Geller, M. J., \& Huchra,
J. P. A slice of the universe. \ApJ\ {\bf 302}, L1-L5
(1986).
\bibitem{Geller89} Geller, M. J., \& Huchra, J. P. Mapping the
universe. {\it Science} {\bf 246}, 897-903
(1989).
\bibitem{deLapparent88} de Lapparent, V., Geller, M. J., \& Huchra,
J. P. The mean density and two-point correlation function for
the CfA redshift survey slices. \ApJ\ {\bf 332}, 44-56
(1988).
\bibitem{Vogeley94} Vogeley, M. S., Park, C., Geller, M. J., Huchra, J. P., \&
Gott, J. R. Topological analysis of the CfA redshift
survey. \ApJ\ {\bf 420}, 525-544
(1994).
\bibitem{Shectman96} Shectman, S.A., Landy, S.D., Oemler, A., Tucker, D.L., Lin,
H., Kirshner, R.P., \& Schechter, P.L. The Las Campanas redshift
survey, \ApJ\ {\bf 470}, 172-188
(1996).
\bibitem{Kormendy87} Kormendy, J., \& Knapp, G., editors {\it Dark Matter in the
Universe} (Dordrecht: Reidel)
(1987).
\bibitem{SW} Strauss, M. A., \& Willick, J. A. The
density and peculiar velocity fields of nearby galaxies. {\it Physics
Reports} {\bf 261}, 271-431
(1995).
\bibitem{Fisher95} Fisher, K. B., Huchra, J. P., Davis, M., Strauss, M. A.,
Yahil, A., \& Schlegel, D. The \iras\ 1.2 Jy survey:
redshift data. \ApJS\ {\bf 100}, 69-103
(1995).
\bibitem{Coles96} Coles, P., \& Lucchin, F. {\it Cosmology: The
Origin and Evolution of Cosmic Structure} (New York: Wiley) 
(1995).
\bibitem{Padmanabhan95} Padmanabhan, T. {\it Structure Formation in the Universe}
(Cambridge: Cambridge University Press)
(1993).
\bibitem{KolbTurner90} Kolb, E. W., \& Turner, M. S. {\it The Early Universe}
(Redwood City: Addison-Wesley)
(1990).
\bibitem{Partridge96} Partridge, R.B. {\it 3K: the Cosmic Microwave
Background Radiation} (Cambridge: Cambridge University Press)
(1994).
\bibitem{White94} White, M., Scott, D., \& Silk, J. Anisotropies
in the cosmic microwave background. \ARAA\ {\bf 32}, 319-370
(1994).
\bibitem{Bertschinger96} Bertschinger. E. Cosmological
dynamics. In {\it Cosmology and
Large-Scale Structure}, ed.~R. Shaeffer \etal\ (Amsterdam: Elsevier Science), 273-347
(1996).
\bibitem{Peebles82} Peebles, P. J. E. Large-scale background
temperature and mass fluctuations due to scale-invariant primeval
perturbations. \ApJ\ {\bf 263}, L1-L4
(1982).
\bibitem{Peacock97} Peacock, J.A. The evolution of galaxy
clustering. \MNRAS\ {\bf 284}, 885-898
(1997).

\bibitem{Fisher93} Fisher, K. B., Davis, M., Strauss, M. A., Yahil, A., \&
Huchra, J. P. The power spectrum of \iras\ galaxies. \ApJ\ {\bf 402}, 42-57
(1993).
\bibitem{Ostriker93} Ostriker, J.P. Astronomical tests of the
cold dark matter scenario. \ARAA\ {\bf 31}, 689-716
(1993).
\bibitem{Kent82} Kent, S.M., \& Gunn, J.E. The dynamics of rich
clusters of galaxies. I - The Coma cluster. \AJ\ {\bf 87}, 945-971
(1982).
\bibitem{Kaiser87} Kaiser, N. Clustering in real space
and in redshift space. \MNRAS\ {\bf 227}, 1-21
(1987).
\bibitem{Hamilton98} Hamilton, A.J.S. Linear redshift
distortions: A review. in {\it Ringberg Workshop on
Large-Scale Structure}, ed. D. Hamilton (Kluwer, Amsterdam), 185-276
(1998).
\bibitem{Kaiser84} Kaiser, N. On the spatial correlations of
Abell clusters. \ApJ\ {\bf 284}, L9-L12
(1984).
\bibitem{DekelRees87} Dekel, A., \& Rees, M. J. Physical
mechanisms for biased galaxy formation. {\it Nature} {\bf 326}, 455-462
(1987).
\bibitem{Kauffman98} Kauffmann, G., Nusser, A., \& Steinmetz,
M. Galaxy formation and large-scale bias. \MNRAS\ {\bf 286}, 795-811 
(1997).
\bibitem{DekelLahav98} Dekel, A., \& Lahav, O. Stochastic
nonlinear galaxy biasing. Preprint, astro-ph/9806193
(1998).
\bibitem{Dressler80} Dressler, A. Galaxy morphology in rich
clusters - Implications for the formation and evolution of
galaxies. \ApJ\ {\bf 236}, 351-365
(1980).
\bibitem{Huchra91} Huchra J. P., Geller, M. J., de Lapparent, V., \& Corwin, H. G. 
The CfA redshift survey - Data for the NGP $+30^\circ$
zone. \ApJS\ {\bf 72}, 433-470 
(1990).
\bibitem{Davis76} Davis, M., \& Geller, M.J. Galaxy correlations
as a function of morphological type. \ApJ\ {\bf 208}, 13-19
(1976).
\bibitem{Davis88} Davis, M., Meiksin, A., Strauss, M. A., da Costa, N., \&
Yahil, A. On the universality of the galaxy two-point
correlation function. \ApJ\ {\bf 333}, L9-L12
(1988).
\bibitem{Guzzo97} Guzzo, L., Strauss, M.A., Fisher, K.B.,
Giovanelli, R., and Haynes, M.  Redshift--space distortions and
the real--space clustering of different galaxy types. \ApJ\ {\bf 489}, 37-48
(1997).
\bibitem{Juszkiewicz93} Juszkiewicz, R., Bouchet, F.R., \& Colombi,
S. Skewness induced by gravity. \ApJ\ {\bf 412}, L9-12
(1993).
\bibitem{Spergel93} Spergel, D.N., \& Turok, N.G. Textures and
cosmic structure. {\it Sci.~Am.} {\bf 266}, 52-59
(1992).
\bibitem{Gott86} Gott, J. R., Dickinson, M., \& Melott, A. L.  The
sponge-like topology of large-scale structure in the universe. \ApJ\
{\bf 306}, 341-357 
(1986).
\bibitem{Bouchet93} Bouchet, F. R., Strauss, M., Davis, M., Fisher, K. B.,
Yahil, A., \& Huchra, J. P. Moments of the counts
distribution in the 1.2 jy \iras\ galaxy sample. \ApJ\ {\bf 417}, 36-53
(1993).
\bibitem{KimStrauss98} Kim, R.S.J., \& Strauss, M.A. Measuring
high-order moments of the galaxy distribution from counts in cells --
the edgeworth expansion. \ApJ\ {\bf 493}, 39-51
(1998).
\bibitem{Gaztanaga95} Gazta\~naga, E. High-order galaxy
correlation functions in the APM galaxy survey. \MNRAS\ {\bf 268}, 913-924
(1994).
\bibitem{Strauss92} Strauss, M. A., Yahil, A., Davis, M., Huchra, J. P., Fisher,
K. B. A redshift survey of \iras\ galaxies V: The acceleration on the local 
group. \ApJ\ {\bf 397}, 395-419
(1992).
\bibitem{Teerikorpi97} Teerikorpi, P. Observational
selection bias affecting the determination of the extragalactic
distance scale. \ARAA\ {\bf 35}, 101-136
(1997).
\bibitem{Willick97} Willick, J. A., Courteau, S., Faber, S. M., Burstein, D.,
Dekel, A., \& Strauss, M. A. Homogeneous
velocity-distance data for peculiar velocity analysis. III. The Mark
III catalog of galaxy peculiar velocities. \ApJS\ {\bf 109}, 333-366
(1997).
\bibitem{daCosta97} da Costa, L.N., Nusser, A., Freudling, W., Giovanelli,
R., Haynes, M.P., Salzer, J.J., \& Wegner, G. Comparison of the
SFI peculiar velocities with the \iras\ 1.2 Jy gravity
field. Preprint, astro-ph/9707299
(1997).
\bibitem{Shaya95} Shaya, E. J., Peebles, P. J. E., \& Tully,
R. B. Action principle solutions for galaxy motions within 3000
kilometers per second. \ApJ\ {\bf 454}, 15-31
(1995).
\bibitem{DNW} Davis, M., Nusser, A., \& Willick, J. A. Comparison of
velocity and gravity fields: the Mark III Tully-Fisher 
catalog versus the \iras\ 1.2 Jy survey. \ApJ\ {\bf 473}, 22-41
(1996).
\bibitem{VELMOD2} Willick, J.A., \& Strauss, M.A. Maximum-likelihood
comparison of Tully-Fisher and redshift data. ii. results from an
expanded sample. \ApJ\ in press (astro-ph/9801307) 
(1998).
\bibitem{Dekel94} Dekel, A. Dynamics of cosmic flows. \ARAA\
{\bf 32}, 371-418
(1994).
\bibitem{Nusser91} Nusser, A., Dekel, A., Bertschinger, E., \& Blumenthal, G. R.,
Cosmological velocity-density relation in the quasi-linear
regime. \ApJ\ {\bf 379}, 6-18
(1991).
\bibitem{Sigad98} Sigad, Y., Eldar, A., Dekel, A., Strauss, M.A.,
\& Yahil, A. \iras\ versus POTENT density fields on large
scales: biasing and Omega. \ApJ\ {\bf 495}, 516-532
(1998).
\bibitem{Dekel97} Dekel, A., Burstein, D., \& White, S.D.M. Measuring
Omega. in {\it Critical Dialogues in Cosmology}, edited by 
N. Turok (Singapore: World Scientific), 175-192
(1997).
\bibitem{Bahcall98} Bahcall, N.A., Fan, X., \& Cen, R. Constraining
$\Omega$ with cluster evolution. \ApJ\ {\bf 485}, L53-L56 
(1997).
\bibitem{GunnWeinberg95} Gunn, J. E., \& Weinberg, D. H. The
Sloan Digital Sky Survey. In {\it
Wide-Field Spectroscopy and the Distant Universe}, ed.\ S. J. Maddox and A.
Arag\'on-Salamanca (Singapore: World Scientific), 3-14
(1995).
\bibitem{Knapp97} Knapp, J.R. Mining the Heavens: The Sloan
Digital Sky Survey, {\it Sky \& Tel.} {\bf 94}, 40-48
(1997).
\bibitem{Colless} Colless, M. First results from the 2dF galaxy
redshift survey. {\it Phil.~Trans.~R.~Soc.~Lond.~A}, in
press (astro-ph/9804079)
(1998).
\bibitem{Wang98} Wang, Y., Spergel, D.N., \& Strauss, M.A. Cosmology
in the Next Millennium: Combining MAP and SDSS data to constrain
inflationary models. Preprint, astro-ph/9802231
(1998).
\bibitem{Efstathiou91b} Efstathiou, G., Bernstein, G., Katz, N., Tyson, J. A., \&
Guhathakurta, P. The clustering of faint galaxies, \ApJ\ {\bf 380}, L47-L50
(1991).
\bibitem{Infante94} Infante, L., \& Pritchet, C.J. The
clustering properties of faint galaxies. \ApJ\ {\bf 439}, 565-583
(1995).
\bibitem{Lauer98} Postman, M., Lauer, T.R., Szapudi, I., \& 
Oegerle, W. Clustering at high redshift: precise constraints
from a deep, wide area survey.  \ApJ\ in press (astro-ph/9804141)
(1998).
\bibitem{Hogg98} Hogg, D.W. \etal\ A blind test of photometric
redshift prediction. \AJ\ {\bf 115}, 1418-1422
(1998).
\bibitem{Connolly98} Connolly, A.J., Szalay, A.S., \& Brunner,
R.J. Evolution of the angular correlation function. \ApJ\ {\bf 499}, L125-129
(1998).
\bibitem{CFRS} Lilly, S.J., Le F\`evre, O., Crampton, D., Hammer, F., \&
Tresse, L. The Canada-France Redshift Survey. I. Introduction to
the survey, photometric catalogs, and surface brightness selection
effects. \ApJ\ {\bf 455}, 50-59
(1995).
\bibitem{CNOC2} Yee, H.K.C., Ellingson, E., \& Carlberg, R.G. The CNOC
cluster redshift survey catalogs. I. observational strategy 
and data reduction techniques. \ApJS\ {\bf 102}, 269-287
(1996).
\bibitem{Steidel96} Steidel, C.C., Giavalisco, M., Pettini, M.,
Dickinson, M., \& Adelberger, K.L. Spectroscopic confirmation of
a population of normal star-forming galaxies at redshifts $z >
3$. \ApJ\ {\bf 462}, L17-L20
(1996).
\bibitem{Steidel98}  Giavalisco, M., Steidel, C.C., Adelberger, K.L.,
Dickinson, M.E., Pettini, M., \& Kellogg, M. The angular
clustering of lyman-break galaxies at redshift $z=3$. \ApJ\ in press
(astro-ph/9802318) 
(1998).
\bibitem{Fry96} Fry, J.N. The Evolution of Bias. \ApJ\ {\bf 461}, L65-L68
(1996).
\bibitem{TegmarkPeebles98} Tegmark, M. \& Peebles, P.J.E. The
time evolution of bias. \ApJ\ {\bf 500}, L79-L82
(1998).
\bibitem{BigTelescopes} Article on Big Telescopes, this issue
\bibitem{Rauch98} Rauch, M. The Lyman alpha forest in the spectra
of QSOs. \ARAA\ in press (astro-ph/9806286)
(1998).
\bibitem{Croft98} Croft, R. A. C., Weinberg, D. H., Katz, N., \&
Hernquist, L. Recovery of the power spectrum of mass
fluctuations from observations of the Ly$\alpha$ forest. \ApJ\ {\bf 495}, 44-62
(1998).
\bibitem{Bahcall88} Bahcall, N. Large-scale structure in the
universe indicated by galaxy clusters. \ARAA\ {\bf 26}, 631-686
(1998).
\bibitem{Rosati98} Rosati, P., Della Ceca, R., Norman, C., \&
Giacconi, R. The ROSAT deep cluster survey: the X-ray luminosity
function out to $z=0.8$. \ApJ\ {\bf 492}, L21-L24 (1998). 
\end{thebibliography}
\end{document}